\documentclass{article}
\usepackage{spconf,amsmath,graphicx}
\usepackage{enumitem}
\usepackage{makecell}
\usepackage{booktabs}
\usepackage{microtype}
\usepackage{caption}
\usepackage{amssymb}
\usepackage{amsmath}
\usepackage{scrlfile}
\usepackage{etoolbox}
\usepackage{multirow}
\usepackage[hidelinks]{hyperref}

\title{Distilling HuBERT with LSTMs via Decoupled Knowledge Distillation}
\name{Danilo de Oliveira, Timo Gerkmann\thanks{Funded by the Federal Ministry of Education and Research (BMBF) and the Free and Hanseatic City of Hamburg under the Excellence Strategy of the Federal Government and the Länder}}
\address{Signal Processing (SP), Universit\"at Hamburg, Hamburg, Germany}%

\begin{document}
\maketitle
\begin{abstract}
Much research effort is being applied to the task of compressing the knowledge of self-supervised models, which are powerful, yet large and memory consuming. In this work, we show that the original method of knowledge distillation (and its more recently proposed extension, decoupled knowledge distillation) can be applied to the task of distilling HuBERT. In contrast to methods that focus on distilling internal features, this allows for more freedom in the network architecture of the compressed model. We thus propose to distill HuBERT's Transformer layers into an LSTM-based distilled model that reduces the number of parameters even below DistilHuBERT and at the same time shows improved performance in automatic speech recognition. %
\end{abstract}
\section{Introduction}
\label{sec:intro}

Self-supervised learning (SSL) has become ubiquitous in the deep learning world. Its unsupervised pre-training stage allows for crafting very powerful and general representations, which can then be leveraged in a supervised fine-tuning stage for specific downstream tasks.
However, a well-known drawback of these models is their size. As an example, the HuBERT model in its smallest version (\texttt{BASE}) contains $\sim$95 million parameters \cite{hsu2021hubert}. In this context, compressing the knowledge acquired by large networks or ensembles of networks into smaller ones is crucial in order to make them easier to use in real-world applications and more accessible for experimentation in both industry and academia \cite{lu2017knowledge, takashima2018investigation}.

Different approaches have been proposed for distilling the knowledge in neural networks. The method of knowledge distillation (KD) proposed in \cite{hinton2015distilling} uses soft probabilities assigned to target classes by a large model (the teacher) as guides for a smaller model (the student). More recently, it has been shown that the loss employed in KD can be decomposed into two parts that are coupled and dependent on the teacher's confidence in the target class \cite{zhao2022decoupled}. By decoupling the two terms and assigning hyper-parameters to them, the distillation process can be adjusted to improve knowledge transfer and subsequently performance. Knowledge distillation has been researched extensively for compressing speech recognition models \cite{lu2017knowledge, takashima2018investigation}, even SSL-based ones \cite{yang2022knowledge}. However, as these methods depend on soft probabilities to work, their application to general-purpose SSL models is not straightforward.

The frame embeddings produced by SSL models for speech are feature vectors of a continuous nature. Works like DistilHuBERT \cite{chang2022distilhubert} and FitHuBERT \cite{lee2022fithubert} make use of the representations output by intermediate layers of HuBERT (so-called \emph{hints}) as targets guiding the student's training \cite{romero2015fitnets}. These methods, however, require that the architecture of the student network be similar to that of the teacher, since the student is designed to mimic the teacher's inner representations by means of linear projections. Therefore, distilled versions of SSL models will typically re-use the Transformer \cite{vaswani2017attention} encoder architecture of the teacher, only with techniques to prune layers and/or reduce layer dimensions. 

It is interesting to point out that the pre-training of HuBERT is in essence a cluster prediction task that can be framed as classification. Therefore, in this work we argue that distillation methods based on the model's scores for each class (logits) could be used to compress it. As a consequence, by not targeting the teacher's inner representations but its pre-training outputs, we are free to explore different architectures like Long Short-Term Memory (LSTM) \cite{hochreiter1997lstm} modules. 

LSTMs have been extensively used in audio applications and alongside temporal convolutional networks (TCNs) were the standard neural-network method for sequence modeling for years before the introduction of Transformers \cite{graves2013speech, weninger2015speech}, still being able to provide strong performance in speech applications \cite{tesch2023insights}. 
Additionally, although the sequential processing of recurrent neural networks hurts parallelization during training, they do not suffer from quadratic complexity with respect to the sequence length as Transformers do and can be efficiently used in streaming applications \cite{graves2013speech}. We therefore propose in this work an LSTM-based architecture to compress the powerful Transformer features obtained during pre-training and evaluate it in various downstream tasks.

\section{Knowledge Distillation}
\label{sec:kd}

The method of knowledge distillation proposed in \cite{hinton2015distilling} consists of transferring knowledge from a potentially large and cumbersome classification model to a distilled one using a soft target distribution for each sample in a transfer dataset, calculated for each of $C$ classes as
\begin{equation}
    p_c = \frac{\exp(l_c/\tau)}{\sum_{c'=1}^C\exp(l_{c'}/\tau)},
\end{equation}
where $l_c$ represents the logit of class $c$, and $\tau$ is a temperature scaling. The authors argue that soft targets provide much more information than hard labels and show that they are an effective way of communicating patterns discovered by a teacher model, being leveraged as guidance for training smaller models even with less data than the teacher. 

The knowledge distillation loss \cite{hinton2015distilling, zhao2022decoupled} is defined as
\begin{equation}
    \mathcal{L}_\mathrm{KD} = \mathrm{KL}(\mathbf{p}^\mathcal{T}||\mathbf{p}^\mathcal{S})\cdot\tau^2,
\end{equation}
where $\mathrm{KL}(\cdot||\cdot)$ is the Kullback-Leibler divergence, and $\mathbf{p}^\mathcal{T}$ and $\mathbf{p}^\mathcal{S}$ are the softmax-probability scores output by the teacher and the student model, respectively. The final loss is a linear combination of $\mathcal{L}_{KD}$ and the cross-entropy loss between the student's class probabilities and the ground-truth hard labels.

More recently, Decoupled Knowledge Distillation (DKD) \cite{zhao2022decoupled} was proposed. The authors first show that the KD loss can be split into two parts: Target Class Knowledge Distillation (TCKD) and Non-Target Class Knowledge Distillation (NCKD). The first one is argued to transfer the knowledge about the "difficulty" of the training samples, while the latter contains valuable knowledge about the non-target classes. The reformulated loss reads as
\begin{equation}
    \mathcal{L}_{KD} = \underbrace{\mathrm{KL}(\mathbf{b}^\mathcal{T}||\mathbf{b}^\mathcal{S})}_{\mathrm{TCKD}}\cdot\tau^2+(1-p^\mathcal{T})\underbrace{\mathrm{KL}(\mathbf{\hat{p}}^\mathcal{T}||\mathbf{\hat{p}}^\mathcal{S})}_{\mathrm{NCKD}}\cdot\tau^2,
\end{equation}
where $\mathbf{b}\in \mathbb{R}^2$ are the binary probabilities of the target class and all the other non-target classes together, and $\mathbf{\hat{p}}\in \mathbb{R}^{C-1}$ represents the probabilities among non-target classes, discarding the target one. The NCKD term is shown to be coupled and inversely tied to the teacher's confidence on the target output, $p^\mathcal{T}$. This means that a good teacher would have a high $p^\mathcal{T}$, suppressing the NCKD component. As a consequence, the effectiveness of distillation of valuable information about the non-target classes would be limited for well-predicted samples. The authors therefore propose to decouple the two terms, weighting them with hyper-parameters $\alpha$ and $\beta$ instead, resulting in the DKD loss:

\begin{equation}
    \mathcal{L}_{DKD} = \alpha\cdot\mathrm{TCKD} + \beta\cdot\mathrm{NCKD}.
\end{equation}

The authors show that DKD improves on the performance of KD in computer vision tasks without any additional trainable parameters, obtaining performance comparable to that of feature-based distillation methods, which motivates us to apply it to the best of our knowledge for the first time in compressing general-purpose audio SSL models such as HuBERT.

\begingroup
\setlength{\tabcolsep}{4pt} %
\renewcommand{\arraystretch}{1.4} %
\begin{table*}[t]
	\centering 
    \scalebox{0.95}{
    \resizebox{\textwidth}{!}{
     \begin{tabular}{c| l c c c c c c c c c c c c c c}
    \toprule
        \multicolumn{1}{c}{} & & & \multicolumn{5}{c}{Content} & \multicolumn{3}{c}{Speaker} & \multicolumn{3}{c}{Semantics} & Paraling. \\ 
        \cmidrule(lr){4-8} \cmidrule(lr){9-11} \cmidrule(lr){12-14} \cmidrule(lr){15-15}
        \multicolumn{1}{c}{} & & & PR & \multicolumn{2}{c}{ASR (WER)} & KS & QbE & SID & ASV & SD & IC & \multicolumn{2}{c}{SF} & ER \\
        \cmidrule(lr){4-4}\cmidrule(lr){5-6}\cmidrule(lr){7-7}\cmidrule(lr){8-8}\cmidrule(lr){9-9}\cmidrule(lr){10-10}\cmidrule(lr){11-11}\cmidrule(lr){12-12}\cmidrule(lr){13-14}\cmidrule(lr){15-15}
        \multicolumn{1}{c}{} & Upstream & Params (M) & PER$\downarrow$ & w/o LM$\downarrow$ & w/ LM$\downarrow$ & Acc$\uparrow$ & MTWV$\uparrow$ & Acc$\uparrow$ & EER$\downarrow$ & DER$\downarrow$ & Acc$\uparrow$ & F1$\uparrow$ & CER$\downarrow$ & Acc$\uparrow$ \\
		\midrule
        \multirow{3}{*}{\rotatebox[origin=c]{90}{\parbox[t][]{2cm}{\centering SSL\\baselines}}} 
    	& HuBERT \texttt{LARGE} \cite{hsu2021hubert} & $316.6$ & $3.53$ & $3.62$ & $2.94$ & $95.29$ & $0.0353$ & $90.33$ & $5.98$ & $5.75$ & $98.76$ & $89.81$ & $21.76$ & $67.62$ \\
        &HuBERT \texttt{BASE} \cite{hsu2021hubert} & $94.7$ & $5.41$ & $6.42$ & $4.79$ & $96.30$ & $0.0736$ & $81.42$ & $5.11$ & $5.88$ & $98.34$ & $88.53$ & $25.20$ & $64.92$ \\
        &LightHuBERT $\textit{a}_{\texttt{SMALL}}$ \cite{wang22lighthubert} & $94.7\rightarrow27.0$ & $6.60$ & $8.33$ & $6.04$ & $96.07$ & $0.0764$ & $69.70$ & $5.42$ & $5.85$ & $98.23$ & $87.58$ & $26.90$ & $64.12$ \\
        \midrule
        \multirow{3}{*}{\rotatebox[origin=c]{90}{\parbox[t][]{2cm}{\centering Distillation\\baselines}}}  
        &FitHuBERT \cite{lee2022fithubert} & $22.5$ & $13.32$ & $12.09$ & --- & $96.27$ & $0.0489$ & $55.71$ & $8.00$ & $6.84$ & $91.25$ & $84.06$ & $32.46$ & $59.82$ \\ 
        &DistilHuBERT \cite{chang2022distilhubert} & $23.5$ & $16.27$ & $13.37$ & $9.21$ & $95.98$ & $0.0511$ & $73.54$ & $8.55$ & $6.19$ & $94.99$ & $82.57$ & $35.59$ & $63.02$ \\
        &DistilHuBERT* (\texttt{LARGE}) & $23.5$ & $13.12$ & $13.01$ & $9.65$ & $96.36$ & $0.0484$ & $71.62$ & $9.17$ & $5.95$ & $94.86$ & $83.39$ & $32.74$ & $63.97$ \\
        \midrule
        \multirow{4}{*}{\rotatebox[origin=c]{90}{\parbox{2cm}{\centering Proposed\\(D)KD\\models}}} 
        &KD DistilHuBERT & $23.5$ & $14.75$ & $13.38$ & $9.29$ & $96.36$ & $0.0596$ & $59.34$ & $7.18$ & $6.54$ & $95.18$ & $84.04$ & $33.59$ & $61.98$ \\
        &KD LSTM HuBERT & $18.8$ & $9.61$ & $10.69$ & $7.51$ & $97.31$ & $0.0483$ & $55.84$ & $8.55$ & $6.54$ & $97.86$ & $84.83$ & $30.86$ & $62.56$\\
        &DKD LSTM HuBERT ($\beta=1$) & $18.8$ & $8.60$ & $10.75$ & $7.70$ & $97.21$ & $0.0458$ & $56.31$ & $7.94$ & $6.77$ & $96.92$ & $85.26$ & $30.50$ & $62.63$\\
        &DKD LSTM HuBERT ($\beta=4$) & $18.8$ & $8.57$ & $10.64$ & $7.70$ & $96.95$ & $0.0535$ & $56.71$ & $7.80$ & $6.48$ & $96.94$ & $85.81$ & $30.73$ & $61.92$ \\
        \bottomrule
	\end{tabular}}
 }
	\caption{Model evaluation on the SUPERB benchmark. The last row block contains our proposed models distilled via (D)KD.}
	\label{tab:superb}%
\end{table*}
\endgroup

\section{HuBERT}
\label{sec:hubert}

HuBERT is an SSL Transformer-based model pre-trained with a masked prediction loss over clusters obtained by iterative training and clustering of embeddings \cite{hsu2021hubert}. In the first iteration, mel-frequency cepstral coefficients (MFCC) features are used to fit a $k$-means model. The clusters assigned to each frame then act as pseudo-labels for the current iteration of model pre-training. In the second iteration, the outputs of the 6th layer of the first iteration model replace the MFCC features for the pseudo-label computation. The model resulting from this stage is HuBERT \texttt{BASE}. A third iteration, this time with a larger model and more data, produces HuBERT \texttt{LARGE}. 

During the pre-training stage, HuBERT takes as input raw audio, encodes it into framed features via a convolutional feature extractor, masks them randomly and uses a BERT-like encoder \cite{devlin_bert_2019} that leverages contextual information to enrich output representations. Given a masked sequence of frames $\Tilde{X}$, at timestep $t$ the distribution for cluster $c$ is calculated with:
\begin{equation}
    p_f(c|\Tilde{X}, t) = \frac{\mathrm{exp}(\mathrm{sim}(A\cdot o_t, e_c)/\tau)}{\sum_{c'=1}^C \mathrm{exp}(\mathrm{sim}(A\cdot o_t, e_{c'})/\tau)},
\end{equation}
where $o_t$ is the output of the Transformer encoder, $A$ denotes a linear projection, $e_c$ is a learned embedding for cluster $c$ and $\mathrm{sim}(\cdot, \cdot)$ represents the cosine similarity operation between two vectors.

Given an input speech utterance $X$ encoded into $T$ frames, the pre-training loss is the cross-entropy computed over masked timesteps, with $M$ denoting the indices to be masked:
\begin{equation}
    \mathcal{L}_m(f; X,M,Z) = \sum_{t\in M}\log(p_f(z_t|\Tilde{X}, t)),
\end{equation}
where $z_t$ is a $C$-class categorical variable, and $Z$ is a vector containing $[z_1,\dots,z_T]$. This classification loss over softmaxed logits makes it suitable for distillation via KD and DKD.

As well-known existing efforts to distill HuBERT, we mention here three feature-based works. The first one, DistilHubert \cite{chang2022distilhubert}, retains the two first Transformer layers of HuBERT and has projections that are trained to match the representations of deeper layers of the teacher in terms of $L_1$ and cosine distance. The second method is FitHuBERT \cite{lee2022fithubert}, which goes deeper and thinner by using the same number of layers as the teacher, but with a reduced feature size. The layers are also matched to the teachers' via projections and a mean squared error loss. Lastly we mention LightHuBERT \cite{wang22lighthubert}, that has a more sophisticated and demanding training procedure that involves firstly a masked distillation stage of HuBERT into a model with the same size, and a second \emph{once-for-all} stage with random subnets. The model obtains impressive performance even in its $\textit{a}_{\texttt{SMALL}}$ version. However, its training process is much more costly than the other distilled baselines. Our goal is to obtain a model that is easy and fast to experiment with even with reduced computing resources. Due to its simplicity and availability in the \texttt{S3PRL}\footnote{\url{https://github.com/s3prl/s3prl}} framework, we use DistilHuBERT as the main baseline.

\section{Implementation Details}

Due to the availability of the checkpoints for the second iteration of HuBERT \texttt{BASE} and its corresponding $k$-means clustering model, we use it to extract the pseudo-labels generated by HuBERT \texttt{BASE} via the \texttt{fairseq}\footnote{\url{https://github.com/facebookresearch/fairseq}} framework. We then feed them into HuBERT \texttt{LARGE} to obtain the logits assigned to each of 500 clusters. Due to the high computational cost of calculating the cosine similarity between each frame and 500 candidate embeddings, in our distilled models we simplify the training objective to a simple classification projection, as in \cite{shi2022learning}. When using the model in downstream tasks, the projection is discarded.

\begin{figure*}[t!]
    \centering
    \includegraphics{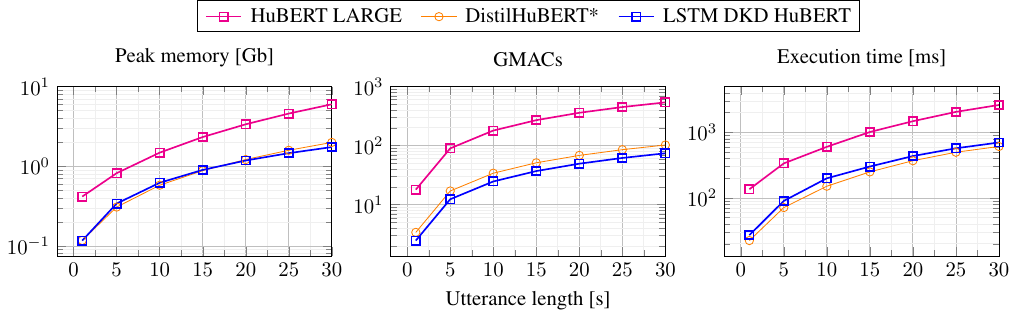}
    \caption{Model profiling on CPU for increasing utterance lengths. \textbf{Left:} Peak memory allocation. \textbf{Middle:} Number of giga multiply-accumulate operations. \textbf{Right:} Execution time, averaged over 10 runs.}
    \label{fig:profiling}%
\end{figure*}

We evaluate the distilled models according to the SUPERB benchmark \cite{yang2021superb}, with 10 tasks grouped into 4 categories: 
\begin{itemize}[leftmargin=*,noitemsep,topsep=0pt,parsep=0pt,partopsep=0pt]
\item\textbf{Content:} Phoneme recognition (PR), automatic speech recognition (ASR), keyword spotting (KS), query-by-example (QbE).
\item\textbf{Speaker:} Speaker identification (SID), automatic speaker verification (ASV), speaker diarization (SD).
\item\textbf{Semantics:} Intent classification (IC), slot filling (SF).
\item\textbf{Paralinguistics:} Emotion recognition (ER).
\end{itemize}

For fine-tuning, we follow the SUPERB methodology of using a weighted sum of the intermediate outputs of the frozen upstream models as input to the downstream models.

Our experiments on DistilHuBERT use the same architecture of two Transformer layers. Since DistilHuBERT originally compresses HuBERT \texttt{BASE} but our teacher model for KD and DKD is HuBERT \texttt{LARGE}, we train a version of DistilHuBERT that compresses the large version, which we denote as DistilHuBERT*. While the original DistilHuBERT uses 3 projections to match its last layer to layers 4, 8 and 12 of HuBERT \texttt{BASE}, we employ 3 extra projections to accommodate for the additional 12 layers of the \texttt{LARGE} model, therefore mapping to layers 4, 8, 12, 16, 20 and 24.

Works such as \cite{romero2015fitnets} focus on the importance of depth for successful representation learning, and it has been shown that different types of information are encoded at different depth levels in SSL models \cite{pasad_layer-wise_2021, deoliveira23_interspeech}. Motivated by \cite{ashihara2022deep} and profiting from the reduced numbers of parameters of LSTM layers compared to HuBERT's Transformer blocks, we try to find a balance between depth and width in our recurrent model: it uses 4 bidirectional LSTM layers with hidden size of 384. This makes the concatenated output features match the dimension of DistilHuBERT, keeping the same reduced footprint. As a consequence, we cannot initialize the two student's layers with the weights of the teachers', which have a feature size of 1024. Nevertheless, it is shown that the initialization with the teacher's weights has only a very minor impact on downstream performance \cite{chang2022distilhubert}. We use the same convolutional feature encoder architecture as the teacher model.

We follow the general distillation framework of DistilHuBERT, using 960 hours of LibriSpeech training data to train the student model for 200k steps, a learning rate of $2 \cdot 10^{-4}$ with a batch size of 32 and a linear warm-up stage for the first 14k updates followed by a linear decay. The cluster assignments are used as the ground truth labels for the cross-entropy loss. The temperature hyper-parameter is set as $\tau=1$. In our KD experiments, we assign equal weights to the cross-entropy classification loss and the KD loss. In the DKD experiments, $\alpha$ is kept at 1 and we explore $\beta=\{1,4\}$.

\section{Results and Discussion}
\label{sec:results}

Table~\ref{tab:superb} shows the results on the 10 tasks of the SUPERB benchmark of the baselines and proposed models. The HuBERT teacher baselines unsurprisingly obtain the best performance across tasks, followed closely by LightHuBERT. The latter involves training a large \emph{once-for-all} Transformer and a second stage of training subnets, resulting in a large network containing smaller subsets of parameters (94.7M$\rightarrow$27.0M for $\textit{a}_{\texttt{SMALL}}$). Due to the larger computational resources required for this process, we place LightHuBERT alongside HuBERT and compare our proposed models rather to Distil- and FitHuBERT.

Firstly analyzing the versions that share the same architecture of DistilHuBERT*, we can see that performance across tasks is generally not changed drastically when using (D)KD, with the exception of speaker identification, which has a significant downgrade in performance. Curiously, the performance of the conceptually related ASV task is improved from DistilHuBERT. In general, the DistilHuBERT architecture distilled via KD and DKD behaves across tasks similarly to FitHuBERT. This could be explained by the fact that even though FitHuBERT uses every layer of the teacher as hints, much more weight is given to the last layer, on which the logit computation is based. Future studies should try to address this performance gap between SID and the other tasks, focusing on global acoustic-related features from earlier layers.

Making use of the freedom acquired from using (D)KD, we experiment with an RNN-based distilled model. Interestingly, our models distilled with LSTMs bring significant improvements to the phoneme recognition and ASR tasks, while using fewer parameters. SF and IC also benefit from the recurrent architecture, while QbE and ASV seem to lose performance if compared with the Transformer-based KD model, although still comparable to Distil- and FitHuBERT. Adding DKD over KD managed to improve scores on PR, ASV and QbE tasks at $\beta=4$. Future studies should address the parametrization of $\alpha$ and $\beta$ for potential further improvements.

Similarly to the masked prediction objective from HuBERT's pre-training, the KD objective focuses on the identification of pseudo-labels, which naturally privileges the linguistic content tasks. It is worth remembering that HuBERT was conceived with a focus on ASR. This objective in combination with the sequential processing of LSTMs result in a boost in ASR and PR performance in comparison to Distil- and FitHuBERT. 

In order to validate that our proposed LSTM model is still resource-efficient and fast, we profile it on memory and execution time with the PyTorch profiler\footnote{\url{https://pytorch.org/docs/stable/autograd.html\#profiler}}, also recording the number of giga multiply-accumulate operations (GMACs) with DeepSpeed\footnote{\url{https://github.com/microsoft/DeepSpeed}}. The results are presented in Figure~\ref{fig:profiling}. The reduced number of layers of DistilHuBERT makes it slightly faster than the LSTM model. The latter, in its turn, involves fewer computations and has a memory allocation that scales linearly with length. Overall, in the analyzed range the distilled models have a similar footprint across all aspects, both providing drastically reduced training times in comparison to the teacher model, making them practical for experimenting both in pre-training and fine-tuning stages.

\section{Conclusions}

Efforts in compressing large Transformer-based audio SSL models so far have mostly focused on feature-based distillation. In this work, we propose the use of the original KD method and its more recent decoupled variant in order to distill HuBERT. This method gives us more flexibility with the choice of architecture, so we propose an LSTM-based model which significantly outperforms DistilHuBERT and FitHuBERT in ASR and phoneme recognition while containing fewer parameters. The use of (D)KD opens up new possibilities in architectural choices, which will be explored in future work.

\clearpage

\bibliographystyle{IEEEtran}
\bibliography{references}

\end{document}